%% file: main.tex
\pdfoutput=1
\documentclass[a4paper, amsfonts, amssymb, amsmath, reprint, showkeys, nofootinbib, twoside,superscriptaddress ]{revtex4-1}
\usepackage[english]{babel}
\usepackage[utf8]{inputenc}
\usepackage[colorinlistoftodos, color=green!40, prependcaption]{todonotes}
\usepackage{array}
\input{preamble}

\usepackage{hyperref}
\hypersetup{colorlinks,allcolors=black}
\newcommand{\Alx}{Al$_\text{1-x}$Sc$_\text{x}$N }
\newcommand{\Alxs}{Al$_\text{1-x}$Sc$_\text{x}$N }
\newcommand{\AlBx}{Al$_\text{1-x}$B$_\text{x}$N }
\newcommand{\AlBxs}{Al$_\text{1-x}$B$_\text{x}$N }
\newcommand{\AlBfix}{Al$_\text{0.61}$B$_\text{0.16}$Sc$_\text{0.21}$N }
\newcommand{\AlScBx}{Al$_\text{1-x-y}$B$_\text{x}$Sc$_\text{y}$N }
\newcommand{\AlScBxs}{Al$_\text{1-x-y}$B$_\text{x}$Sc$_\text{y}$N }

\newcommand{\AlSc}{Al$_\text{0.75}$Sc$_\text{0.25}$N }

\begin{document}
\title{The effect of boron incorporation on leakage and wake-up in ferroelectric \Alxs}

\author{Maike Gremmel}
\email[Correspondence email address: ]{magr@tf.uni-kiel.de}
\affiliation{Institute of Material Science, Kiel University, Germany}

\author{Chandrashekhar Prakash Savant}
\affiliation{Department of Materials Science and Engineering, Cornell University, United States}
\author{Debaditya Bhattacharya}
\affiliation{School of Electrical and Computer Engineering, Cornell University, United States}

\author{Georg Schönweger}
\affiliation{Institute of Material Science, Kiel University, Germany}
\affiliation{Fraunhofer Institute for Silicon Technology (ISIT), Germany }

\author{Debdeep Jena}
\affiliation{Department of Materials Science and Engineering, Cornell University, United States}
\affiliation{School of Electrical and Computer Engineering, Cornell University, United States}
\affiliation{Kavli Institute at Cornell for Nanoscale Science,Cornell University, United States}

\author{Simon Fichtner}
\affiliation{Institute of Material Science, Kiel University, Germany}
\affiliation{Fraunhofer Institute for Silicon Technology (ISIT), Germany }
\date{\today} 

\begin{abstract}
This study explores the influence of boron incorporation on the structural and electrical properties of ferroelectric Aluminum Scandium Nitride (\Alxs) thin films, focusing on leakage currents, wake-up effects, and imprint behavior. \Alxs films were incorporated with varying boron concentrations and analyzed under different deposition conditions to determine their structural integrity and ferroelectric performance. Key findings include a reduction in leakage currents, non-trivial alterations in bandgap energy as well as an increasing coercive fields with increasing boron content. Films with 6-13 at.\% boron exhibited N-polar growth, while those with 16 at.\% boron showed mixed polarity after deposition, which affected their ferroelectric response during the initial switching cycles - as did the addition of boron itself compared to pure \Alxs. With increasing boron content, wake-up became gradually more pronounced and was strongest for pure \AlBx.

\end{abstract}


\maketitle
\section{Introduction}

In recent years, the quest for innovative materials for microelectronic applications has intensified, driven by the need for reduced power consumption and increased functionality, while at the same time maintaining compatibility with existing technology platforms. Among these materials, piezo- and ferroelectric III-nitride based compounds have emerged as a promising contender, showcasing a plethora of positive attributes that position it as a promising emerging material in the field of microelectronics.\cite{Fichtner2019AlScN:Ferroelectric,Hayden2021FerroelectricityFilms,Wang2023DawnDevices, Uehara2021DemonstrationMethod,Kim2023WurtziteMemory}

The intrinsic characteristics of the wurtzite-type ferroelectrics, such as \Alxs, make the material a compelling choice for numerous applications. Notably, it is lead-free, aligning with contemporary environmental standards and regulations. Its compatibility with complementary metal-oxide-semiconductor (CMOS) technology further enhances its appeal, facilitating seamless integration into existing electronic systems.\cite{Islam2021OnFilms,Fichtner2023Wurtzite-TypeFETs,Wang2023DawnDevices, Mikolajick2021NextApplications,Kim2023WurtziteMemory}

One of the standout features of \Alxs is its high spontaneous polarization and high resistance to spontaneous depolarization, which make the material of interest for ferroelectric random access memory (FRAM)\cite{Fichtner2019AlScN:Ferroelectric}. Moreover, its exceptional temperature stability, good long-term stability and chemical resilience underscore its reliability in demanding environments.\cite{Islam2021OnFilms} 
Furthermore, \Alxs boasts relatively low permittivity, contributing to its energy efficacy in various electronic and electromechanical applications.\cite{Schonweger2022FromTechnology,Fichtner2023Wurtzite-TypeFETs} 

The combination of those properties in \Alxs open avenues for its utilization in cutting-edge technologies such as neuromorphic computing, ferroelectric memories, as well as micro-actuators and -sensors - in particular if they are to be used in harsh environments.\cite{Bez2004Non-volatileMaterials,Wessels2007FerroelectricOptics,Khan2020TheTechnology}

Despite numerous advantageous properties, \Alxs is not exempt from a few inherent drawbacks that can influence its performance and reliability in the aforementioned applications. One notable drawback of \Alxs lies in its susceptibility to leakage due to the high coercive fields ($E_c$) that are required to invert its polarization ($P$).\cite{Chen2023LeakageFilm}\cite{Yang2023StressWafer} This characteristic poses challenges, particularly for applications where the risk of electrical failure and leakage must be carefully managed to ensure device integrity and longevity. During the deposition process and device operation, another challenge arises as one polarization direction can become favored over the other. This leads to imprint, i.e. the coercive field for one polarity having a larger absolute value than for the other, that varies over time as well as wake-up, i.e. an increase in the switching polarization during the initial switching cycles.\cite{Schenk2015ComplexOxide, Zhou2005MechanismsFilms, Grossmann2002TheDependence, Zhu2022WakeUpFilms, Gremmel2024TheAl0.70Sc0.30N}
This phenomenon can significantly complicate the operation of memory devices. 

To address the drawbacks associated with \Alx, one promising approach is the incorporation of different materials through doping/alloying.\cite{Kudrawiec2020BandgapEmitters} Incorporation of high bandgap compounds can potentially increase the bandgap of the alloy, thereby reducing leakage. Boron doping presents a particularly intriguing possibility, as \AlBxs has been shown to be ferroelectric itself and the bandgap of BN ($> 6$ eV) is larger than the bandgap of \Alxs ($~4.7$ eV at x$=0.30$).\cite{Levinshtein2001PropertiesSiGe,Baeumler2019Optical0.41} On the other hand, also the coercive field of \AlBxs is higher than that of \Alx, which complicates switching it with on-chip voltage supplies and \AlBxs  might be more prone to feature wake-up.\cite{Hayden2021FerroelectricityFilms, Zhu2022WakeUpFilms}

This work therefore determines whether the advantages of \Alxs and \AlBxs can be combined by forming a quaternary solid solution of \AlScBx. 
The effect of doping \Alxs films with varying amounts of boron via co-sputtering is studied and complemented by structural and electrical analysis to evaluate film quality, boron content, lattice parameters, and other pertinent characteristics under different deposition conditions. We aim to investigate the impact of boron doping on key properties such as the coercive field, bandgap, leakage as well as the initial imprint and wake-up. 

\section{Materials and Methods}
\AlScBx films of 250 nm thickness with different boron concentrations were prepared on oxidized Si(100) wafers which were covered with a 100 nm thick Pt bottom electrode. Al and Sc were co-sputtered with a constant power ratio of 680 W to 320 W from DC sources using pulsed DC-mode while B was deposited simultaneously from an RF source with a power supply of 0 to 400 W. All depositions were performed with single element targets (3N or better) in an Oerlikon MSQ 200 co-sputter system. 
Different deposition temperatures with a fixed power supply of 400 W applied to the boron target were applied to tune the crystal quality and electrical properties. The specific experimental parameters are stated in the results section of this work.
The boron content was measured using X-ray photoelectron spectroscopy (ThermoScientific NEXSA XPS). The surface was cleaned by Ar+ ion beam etching for 40 seconds and the average of 5 data points on the sample was used. 
For structural investigation, scanning electron microscopy (SEM) (Zeiss Gemini Ultra55) was used to determine the thickness of the films as well as the homogeneity of the deposition (e.g. the presence of abnormally oriented grains, AOGs).\cite{Fichtner2017IdentifyingSystems} X-ray diffraction (XRD) (Rigaku SmartLab instrument, 9V,Cu) $\Theta - 2\Theta$ scans and $\omega$ rocking curves (RCs) were used to determine lattice parameters as well as crystal orientation. The lattice parameter \textit{c} was determined by the 0002 \Alxs peak and \textit{a} calculated with help of \textit{c} from the 0-115 reflection (supplement S2).
For electrical characterizations, an additional Pt top-electrode of 100 nm thickness was deposited ex-situ and structured via ion beam etching. Current density over electric field measurements ($J-E$) as well as capacitance over electric field measurements ($C-E$) were performed using a triangular voltage signal in an aixACCT TF Analyzer 2000 to determine coercive field, permittivity and leakage current respectively. 
The breakdown strength was extracted by increasing the applied electric field in multiple $C-E$ measurements until an electric breakdown was induced. Additionally, the bandgap of the different films was determined via Tauc fit from Ellipsometry (Figure S1).

\section{Results and Discussion}
\subsection{Impact of growth temperature on film properties}

\begin{figure}[h]
\includegraphics[width=8cm]{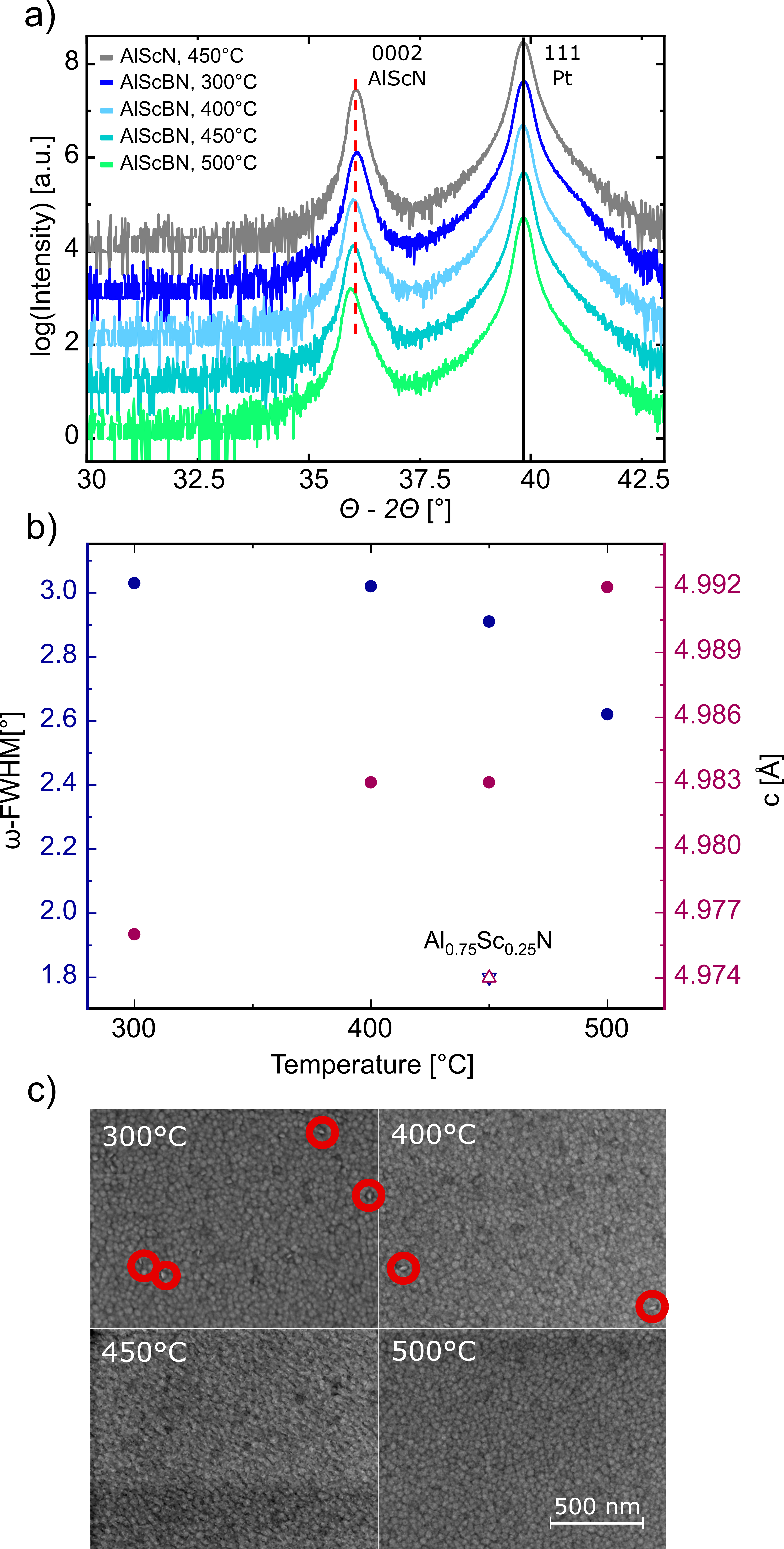}
\centering
\caption{ a)XRD $\Theta - 2\Theta$ scan for \AlBfix 
deposited at different temperatures compared to \AlSc, b) Full width at half maximum of the respective Rocking curves of 0002 reflections and the lattice parameter c.  c)  Respective SEM Images of \AlScBxs films at different deposition temperatures. AOGs are marked by red circles.}
\label{fig:t2t temperature}
\end{figure}

Based on the structure zone model for sputter deposition, one of the main parameters influencing crystal growth is the deposition temperature. Therefore, growth temperatures between 300°C and 500°C were applied, using a fixed power of 400 W applied to the boron target and process gas flows of 7.5 sccm Ar as well as 15 sccm N$_2$  to study the structural and ferroelectric response of the \AlScBxs material. The aim was to determine process settings that result in films with good crystallinity and a uniform polarization direction over the wafer surface. The boron concentration for these \AlScBxs films is determined by XPS to be 16 at.\%. Figure \ref{fig:t2t temperature} a  shows symmetric XRD scans for \AlBfix films grown at different temperatures and that of an \AlSc film. With increasing the deposition temperature, the lattice constant \textit{c} of the \AlBfix films increases by 0.16 pm from 4.976 \AA   at  300°C  to 4.992 \AA  at 500°C, indicating increasingly compressive in-plane stress (Figure \ref{fig:t2t temperature} b). Increasing the deposition temperature improves the XRD rocking curve-FWHM (Full Width at Half Maximum) of the 0002 \AlBfix reflection from 3.0° at 300°C to 2.6° at 500°C, indicating better crystal orientation. While depositions at 300°C and 400°C featured a certain amount of AOGs, 450°C and 500°C result in a homogeneous surface (and uniform grain size) as shown in Figure \ref{fig:t2t temperature} c. The \AlBfix films grown at higher substrate temperature have high crystallinity and fewer AOG.

\begin{figure}[!]
\includegraphics[width=7cm]{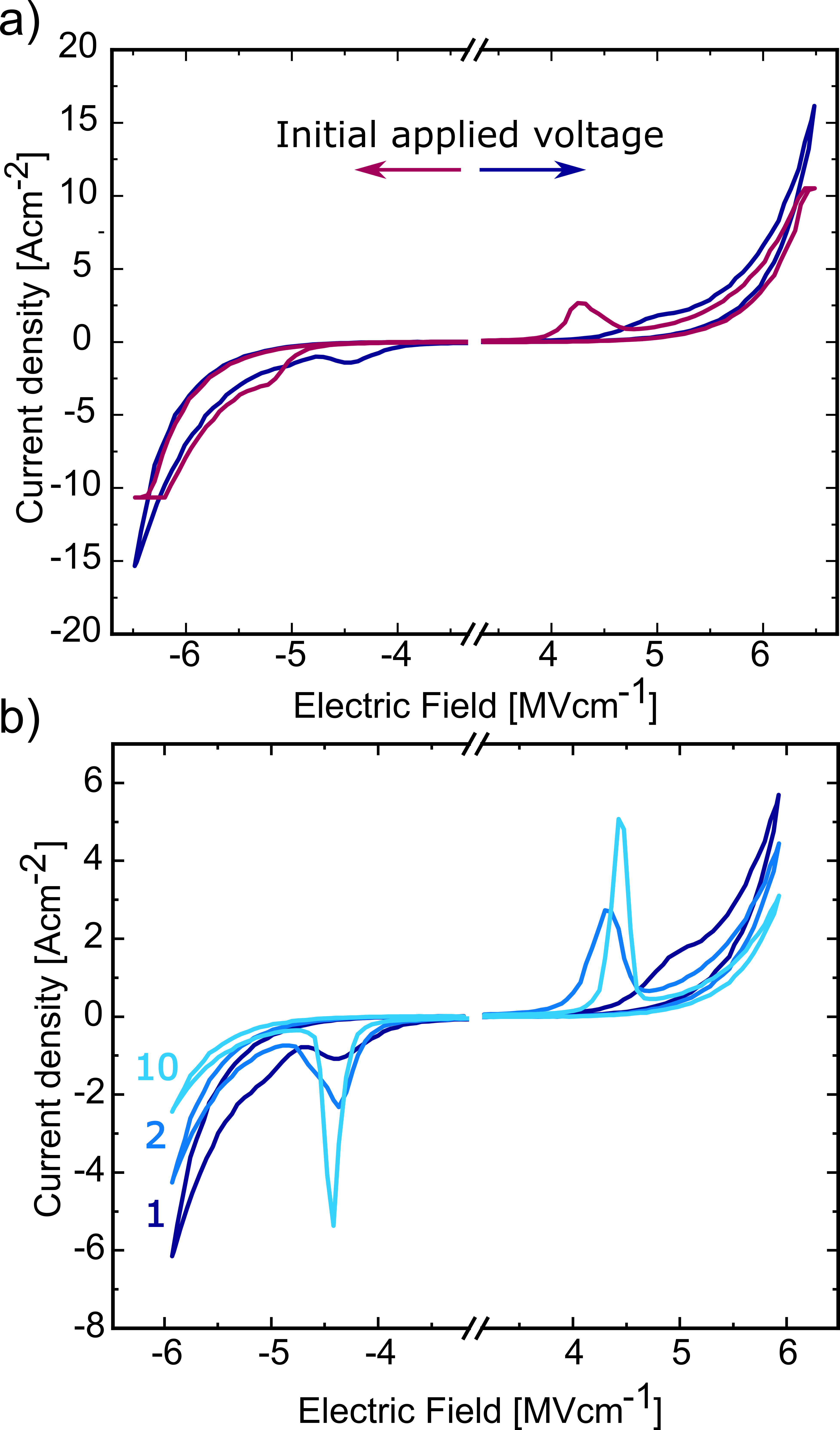}
\centering
\caption{ \AlBfix  deposited at 450°C. a) First cycle on pristine sample for negative voltage train vs positive applied voltage. b) $J-E$ peak formation from first to tenth cycle.}
\label{fig:first cycle}
\end{figure}

All samples exhibit ferroelectric behavior(Figure S3). Textured wurtzite ferroelectrics like \Alxs can feature either nitrogen polarity (N-polarity, polarization pointing downward) or metal polarity (M-polarity, polarization pointing upward). While \Alxs grows exclusively N-polar under the given process parameters, the boron-doped samples show displacement current peaks that result from ferroelectric switching for both electric field directions in the first cycle. Therefore, they feature mixed polarity after growth (figure \ref{fig:first cycle}). A notable change in the shape of the $J-E$ curve is observed between the first and subsequent switching cycles, which manifests in the emergence of a switching current peak. In contrast, the first switching cycle features broad hysteresis areas at substantially larger electric fields. This behavior can be interpreted such that \AlBxs (unlike \Alx) starts to switch very inhomogeneously in terms of the electric field interval under which switching takes place. None the less, it is able to recover the narrow electric field distribution characteristic of wurtzite ferroelectrics after a few cycles. 

While N-polar \Alxs typically features negative imprint, \AlScBxs with mixed polarity shows a more symmetric hysteresis.\cite{Fichtner2019AlScN:Ferroelectric, Gremmel2024TheAl0.70Sc0.30N}

As discussed previously, unipolar films initially lack pre-nucleated domains and/or low energy nuleation sites for the opposite polarity after deposition.\cite{Zhu2022WakeUpFilms, Gremmel2024TheAl0.70Sc0.30N} This asymmetry can facilitate an initial imprint. 
 For mixed-polar samples, M-polar regions likely have a positive imprint, and N-polar regions a negative one. This results in peak splitting/broadening, where one part of the volume switches at lower electric fields while the volume that was stabilized in opposite polarization direction during deposition switches at higher fields. During the first switching of pristine samples, only the volume with the as-grown opposite polarity switches, requiring higher electric fields.

As \Alxs films with similar texture under similar process parameters are purely N-polar in its as-grown state, it can be suspected that the inclusion of boron itself leads to a stronger tendency for mixed-polar growth. 



\subsection{Influence of boron concentration on the structural properties}

Next, the effect of boron concentration was investigated. Since 16 at.\% is already close to the reported stability limit of \AlScBxs, the variation in composition was directed towards less than 16 at.\%.\cite{Hayden2021FerroelectricityFilms}
For these depositions, a temperature of 450°C was chosen, due to the improved 0002 rocking curve FWHM as well as comparability to existing sample sets of \Alxs without boron. To vary the boron content, the power to the RF source was set to 0, 200 and 300 W. Furthermore, one more sample was deposited at 400 W, 7.5 sccm Ar and 10 sccm N$_\text{2}$, which turned out to re-stabilize exclusively N-polar growth at 13 at.\% boron. ´
To compare with literature on \AlBxs, an additional \AlBxs film without Sc was deposited with 1000 W applied to the Al target and 300 W to the boron target. 

\begin{figure}
    \centering
    \includegraphics[width=7cm]{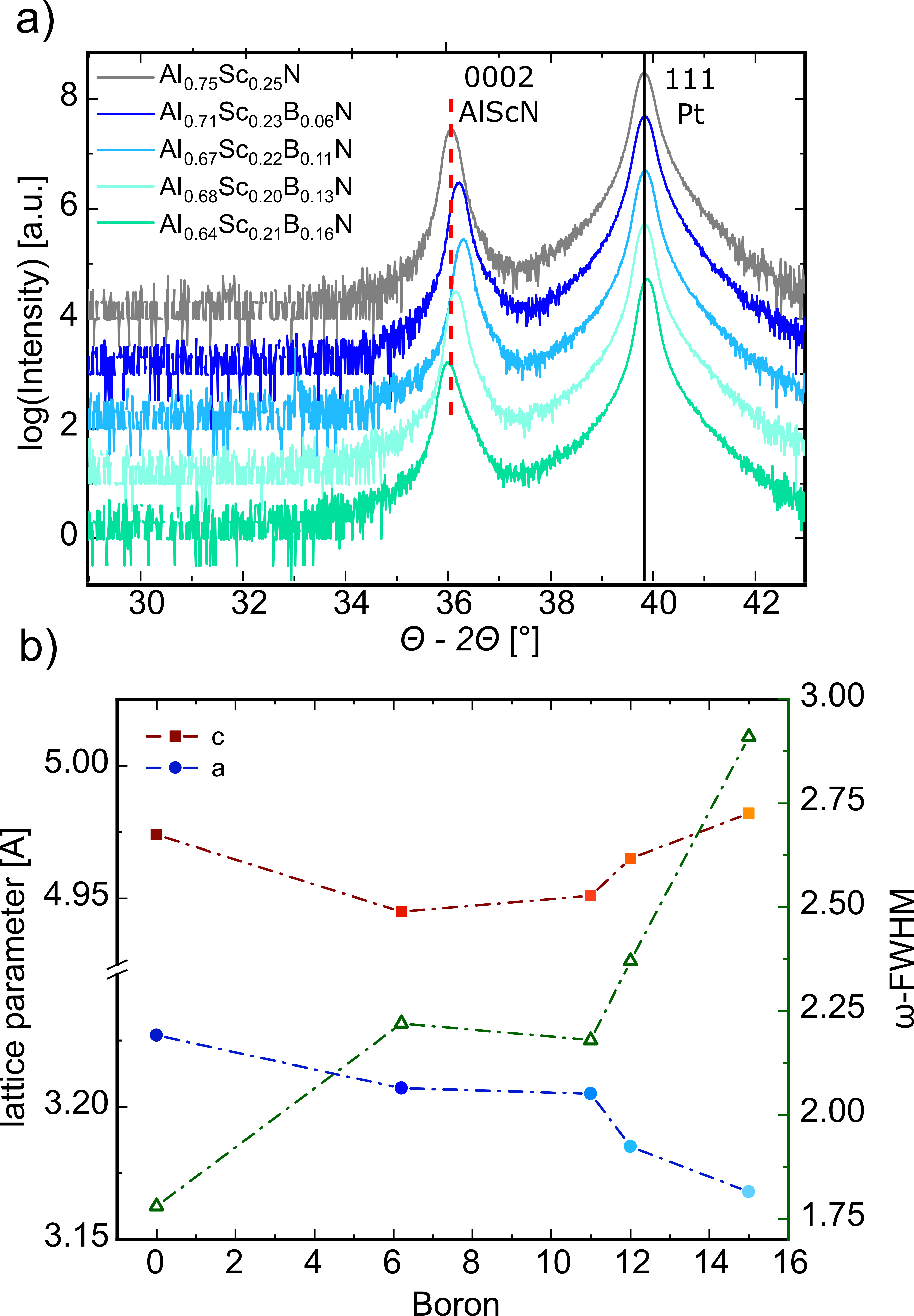}
 \caption{a) XRD $\Theta2\Theta$ scan for different boron concentrations of \AlScBxs and b)lattice parameters \textit{a, c} and FWHM of the corresponding films.}
    \label{fig:XRD boron}
\end{figure}

\begin{figure}
    \centering
    \includegraphics[width=7cm]{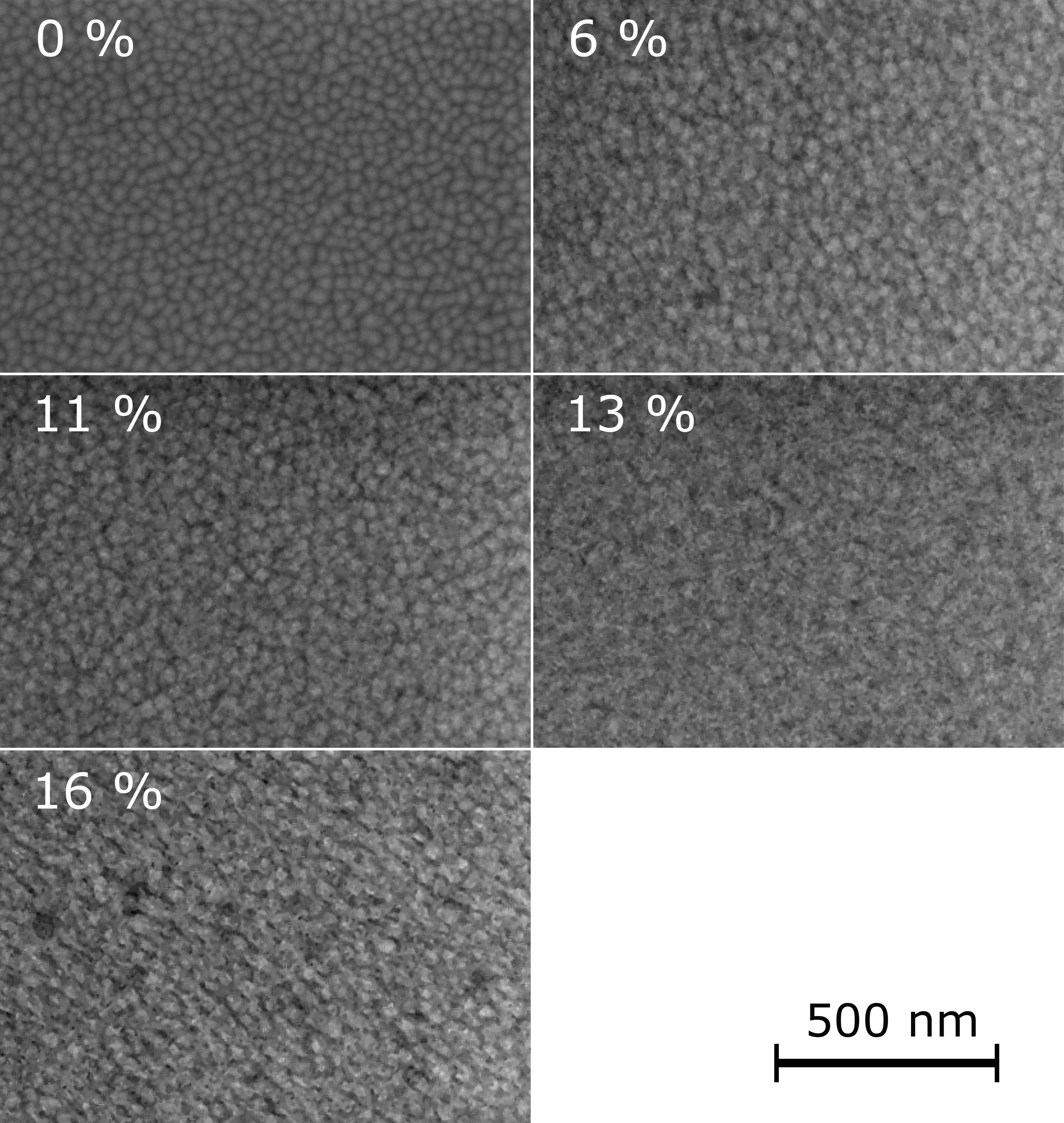}
 \caption{SEM images of the \AlScBxs surface for different B concentrations.}
    \label{fig:SEM boron}
\end{figure}

An overview of all depositions is listed in table \ref{tab:structure} together with the lattice parameters determined via XRD (Figure S2). The atomic ratio Sc/(Al+Sc) determined via XPS remains unchanged at around 25\% for all samples. 
In $\Theta$- 2$\Theta$ scans, an initial decrease in lattice parameter \textit{c} is observed for 6\% and 11\% boron compared to \Alxs, which then increases for 13\% and 16\% B. While \textit{c} for the 13\% \AlScBxs is still smaller than \AlSc, for 16\% it becomes larger (figure \ref{fig:XRD boron}). \textit{a} becomes smaller with increasing B concentration. The FWHM of a rocking curve scan around 0002, as listed in table \ref{tab:structure}, shows a slight decrease in crystal quality towards samples with higher B content. \AlScBxs with 6\%, 11\% 13\% demonstrate similar crystal qualities, as does \AlBx, while the FWHM gets significantly larger for depositions at 400 W, but none the less stays below 3°. The $c/a$ ratio increases upon boron incorporation and is highest for the \AlBfix sample, with 1.576 (table \ref{tab:structure}). This is contrary to observations on \AlBx, where the $c/a$ ratio stays constant at around 1.6.\cite{Hayden2021FerroelectricityFilms}  
The SEM surface images of all films deposited at 450°C are free from AOGs (Figure \ref{fig:SEM boron} b).

\begin{table}[h]
    \centering
    \begin{center}
    \caption{Power applied to the boron target ($P_B$), $_2$ flow, 0002 RC FWHM and lattice constants of the various \AlScBxs samples.}
    \begin{tabular}{m{1.5cm}| m{1cm} m{1cm} m{1cm} m{1cm} m{1cm} m{1cm} }
        $P_\text{B}$ [W]&$N_2$ [sccm]& $B$ [at.\%]& $FWHM$ [°]& $c$ [\AA]&  $a$ [\AA]& $c/a$\\ \hline
         0  &15& 0 & 1.78 & 4.975 & 3.227 & 1.542 \\
        200  &15& 6& 2.22 & 4.945 & 3.207 & 1.542 \\
        300  &15& 11& 2.18 & 4.951 & 3.205 & 1.545 \\
        300,0 (Sc)  &15& --& 2.15 & 5.083 & 3.122 & 1.628 \\
        400 &15& 16& 2.91 & 4.981 & 3.158 & 1.576 \\
        400&10& 13& 2.37 & 4.964 & 3.168& 1.567\end{tabular}
    \label{tab:structure}
    \end{center}
\end{table}

\subsection{Influence of boron concentration on the (ferro-) electric properties}

\begin{figure}
    \centering
    \includegraphics[width=8cm]{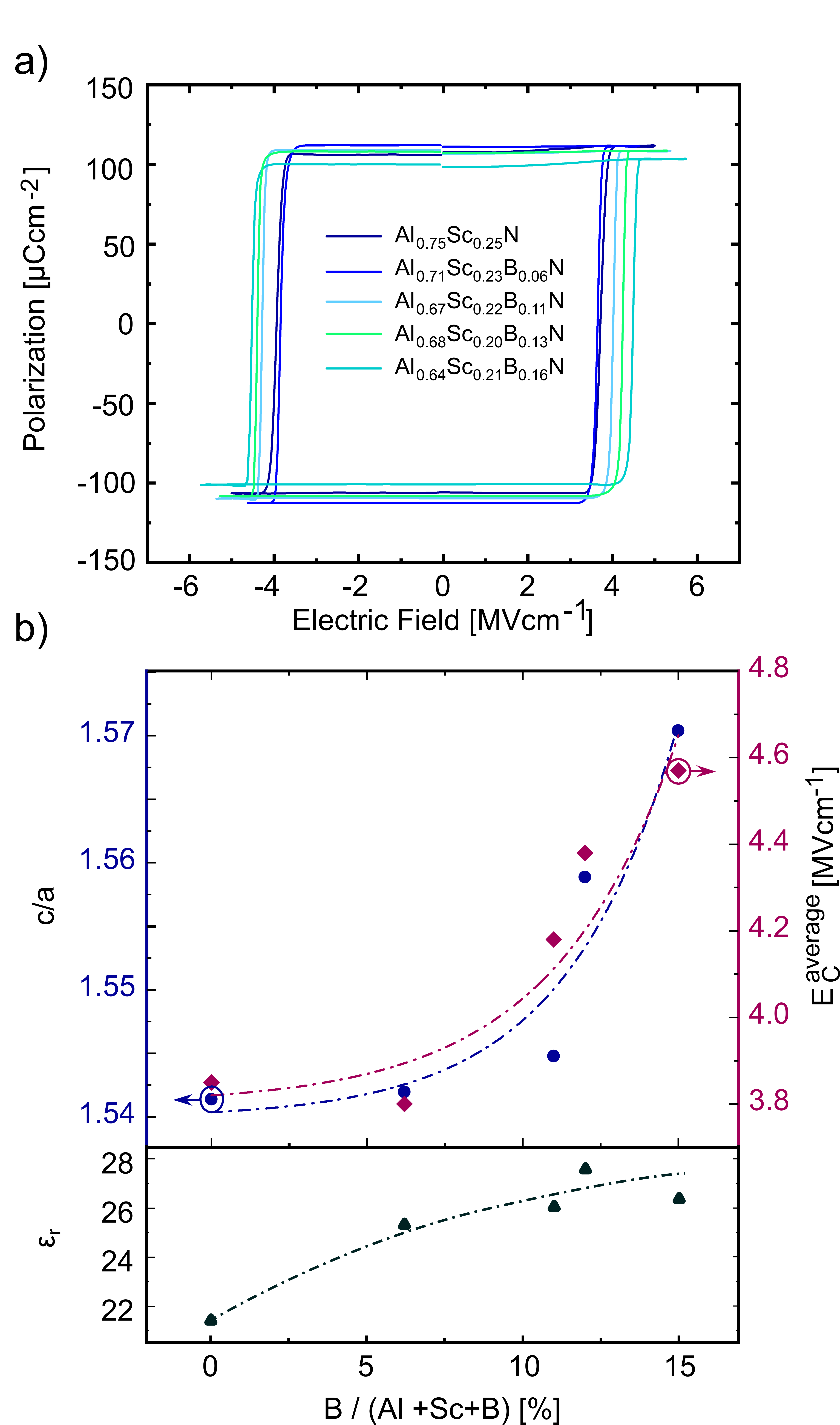}
 \caption{a) $P-E$ hysteresis of \AlScBxs 
 samples and b) $E_c^{average} = (E_{c^+} + E_{c^-})/2$ vs boron concentration as well as $c/a$  and $\varepsilon_r$ vs boron concentration. All $P-E$ loops in a have their leakage current subtracted through peak fitting, as previously described.\cite{Gremmel2024TheAl0.70Sc0.30N}}
    \label{fig:IV-boron conc}
\end{figure}

\begin{figure}
    \centering
    \includegraphics[width=7cm]{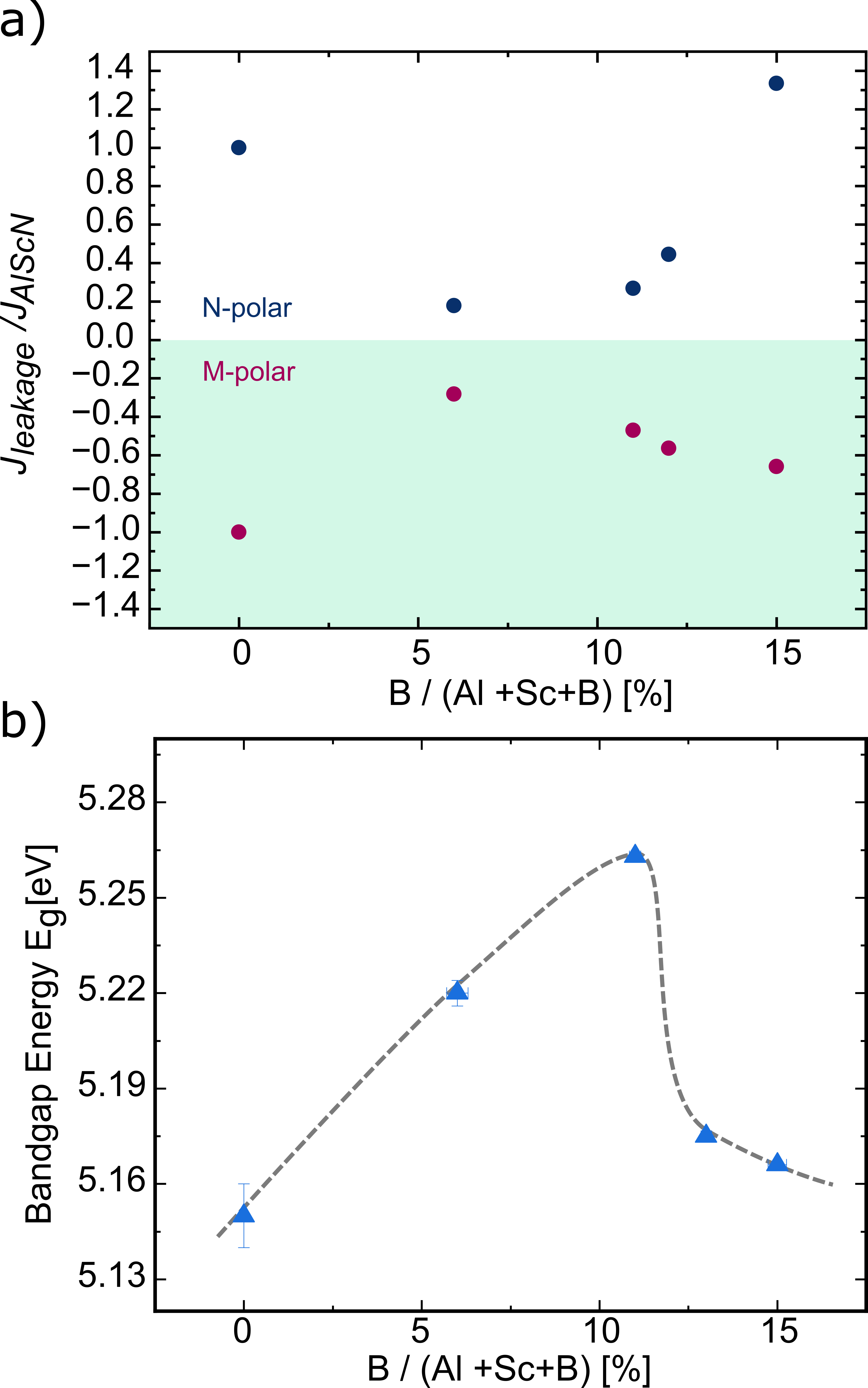}
    \caption{a) Leakage current $J_{leakage}$ normalized to leakage of \AlSc extracted from \textit{J-E} curves that were themselves normalized to their respective $E_c$  and b) $E_g$ vs. boron concentration.}
    \label{fig:leakage}
\end{figure}

Ellipsometry measurements, (Figure S1), show an initial increase in bandgap for increasing boron concentration in the films up to 11\%, before it decreases back to similar values as \AlSc. This is contrary to calculations and experimental data published on \AlBx,\cite{Hayden2021FerroelectricityFilms} where an increase in boron leads to a reduced bandgap $E_g$ - however starting from the 6 eV of pure AlN,\cite{Feneberg2010High-excitationAlN} compared to 5.17 eV of \AlSc from this work. The different behavior can be motivated by the higher bandgap of BN in comparison to ScN. As ScN reduces the bandgap of \AlScBxs, BN apparently counteracts this effect as intended. As can be expected from a material with increased $E_g$, larger boron concentrations do increase the breakdown field $E_{Bf}$ (Figure S6). However, at the same time also $E_c$ (Figure \ref{fig:IV-boron conc}) increases with increasing boron concentration, which is opposite to investigations on \AlBxs - yet again plausible considering the in general higher coercive field of \AlBxs compared to \Alxs. \cite{Hayden2021FerroelectricityFilms} Because of the more rapid increase of $E_c$, the ratio of breakdown voltage to $E_c$ decreases for larger boron concentrations (all values are listed in Table \ref{tab:electrical}). However, there is a significant improvement in leakage currents compared to \Alx, even if the electric field is normalized to the respective $E_c$. This is shown in Figure \ref{fig:leakage}. Up to 13 at.\% boron, the \AlScBx films demonstrate a strong improvement of leakage with 6 at.\% having an optimum of 80\% reduction compared to \AlSc. The 16\% sample shows an improvement of leakage in M-polarity but gets worse for N-polarity. While the 16\% B containing film is mixed polar, all other samples grow exclusively N-polar (Figure S4). Overall, the reduction of leakage is consistent with the observed change in bandgap energy. The samples with higher bandgaps show a reduced leakage, while the \AlScBxs sample with highest boron content and \AlSc feature a similar bandgap and leakage. The $J-E$ curves of which the leakage is extracted is included in the supplement (Figure S5). In spite for the increasing coercive field, the relative permittivity $\varepsilon_r$ is increasing as well, contrary to what is observed for \Alxs.\cite{Fichtner2017IdentifyingSystems}
The relative permittivity of the AlScN film was measured to be 21.4 at 0\%B. The relative permittivity of the AlScBN films increases with increasing the B\% to upto ~26.4 - 27.6 for 13-16\% B. The B incorporation in the AlN is reported to increase its dielectric constant,\cite{Hayden2021FerroelectricityFilms} 
thus an increase in dielectric constant of AlScBN film compared to AlN could be expected. However, with respect to the increasing coercive field with increasing B incorporation, the increase of dielectric constant is unexpected, as a higher $E_c$ usually correlates to lower $\varepsilon_r$ in ferroelectrics.\cite{Mikolajick2021NextApplications}

\begin{table}
    \centering
    \begin{center}
    \caption{Electrical characteristics of \AlScBxs films}
    \begin{tabular}{m{4em} | m{1.5cm} m{1.5cm} m{1.5cm} m{1.5cm}}
        $B $[at.\%]& $E_g$\footnote{Ellipsometry} (eV)& $E_{\text{Bf}}$/$E_c$& $P_r$  \newline [\textmu C cm$^{-2}$]& $\varepsilon_r$ \\ \hline
         0 & 5.17& 2.22& 109 &21.40\\
        6& 5.23& 2.15& 112 & 25.31\\
        11& 5.26& 2.00& 104 &26.04\\
        13& 5.19& 1.74& 108 &27.57\\
        16& 5.16& 1.85 & 106& 26.36\\\end{tabular}
    \label{tab:electrical}
    \end{center}
\end{table}

\begin{figure*} 
    \centering
    \includegraphics[width=17cm]{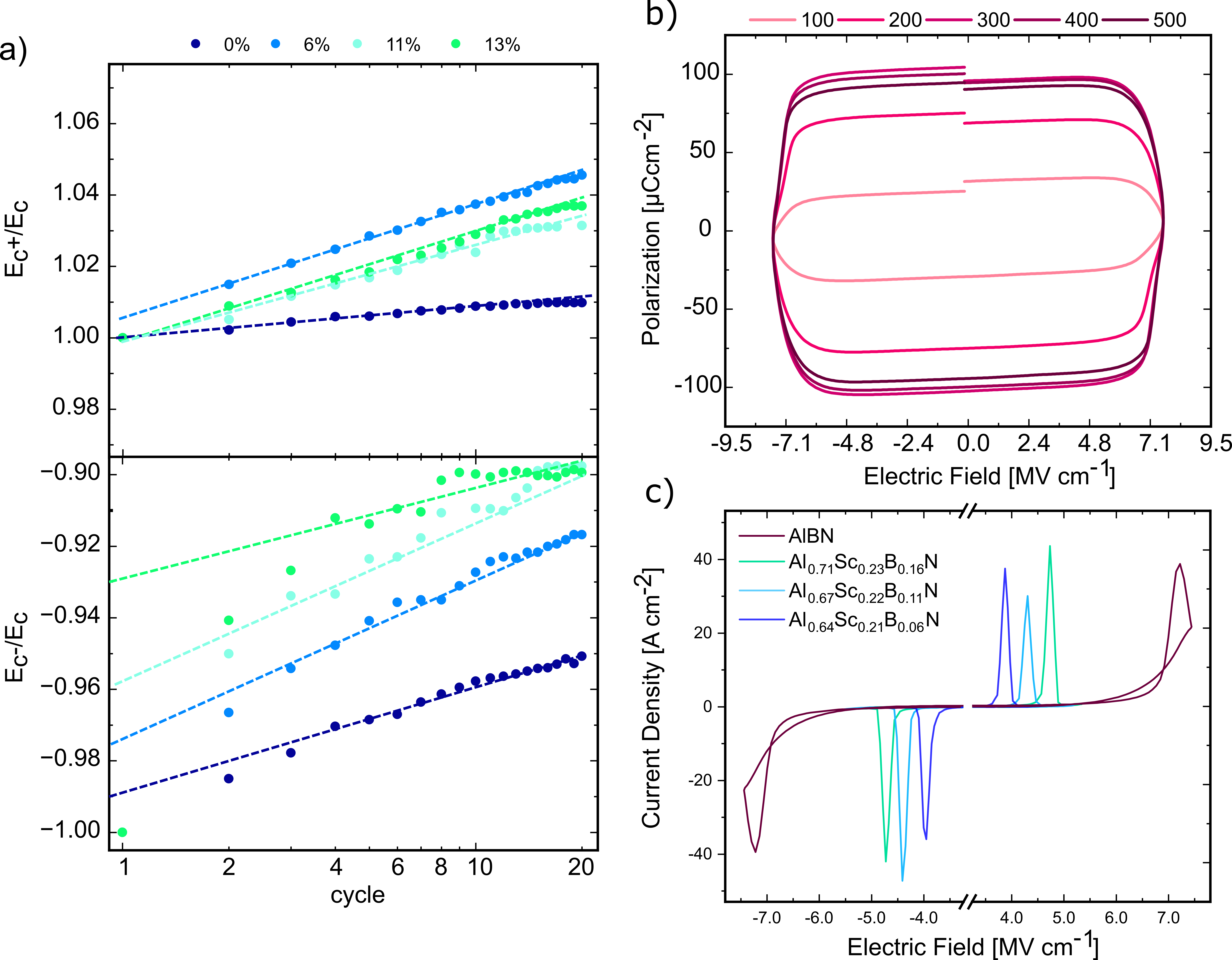}
    \caption{a) $E_c$ of different \AlScBxs films over the number of cycles, b) wake-up process of \AlBxs and c) comparison of $J-E$ curves of different \AlScBxs films with \AlBxs.}
    \label{fig:imprint_wake-up}
\end{figure*}

In terms of remanent polarization, no clear trend towards higher or lower values with increasing boron concentration can be identified, which suggests the value and thus also the spontaneous polarization of the lattice is relatively independent of the investigated composition interval. 
For the pure \AlBxs film, $E_c$ is more than 3 MVcm$^{-1}$ higher than for \AlScBxs deposited at the same target power (Figure \ref{fig:imprint_wake-up}c). For \AlBxs, Zhu et al measured 6 MVcm$^{-1}$ at 100 Hz on epitaxial films at lower B concentration.\cite{Zhu2022WakeUpFilms} To compare both values, the $E_c$ at frequencies between 500 - 2500 Hz were extrapolated to determine $E_c$ at 100 Hz (Figure S7) (Due to leakage, it was not possible to measure $E_c$ directly at lower frequencies). With 7.28 MVcm$^{-1}$ at 100 Hz, the $E_c$ is still significantly higher than for the epitaxial film even though the coercive field of \AlBxs was shown to decrease with higher B content. \cite{Hayden2021FerroelectricityFilms} Furthermore, compared to \Alxs, where the full $P_r$ can easily be switched in the first cycle, a distinct displacement current peak appears in the $J-E$ curves only after several hundred cycles, resulting in the opening of the $P-E$ hysteresis loops. (Figure \ref{fig:imprint_wake-up} b). 

When plotting the normalized coercive fields extracted from the \textit{J-E} measurements against the number of cycles, as shown in Figure \ref{fig:imprint_wake-up} a, it becomes evident that boron significantly impacts imprint and imprint shift during the first few switching cycle. The shift in negative $E_c$ with cycling for \AlSc follows a logarithmic relationship, with all measurements aligning linearly on a logarithmic scale. With increasing boron content, the data deviates more from this logarithmic relationship in particular for the initial three switching cycles. For 13\% boron, $E_c$ shifts by 6\% from the first to the second measurement while the shift of \AlSc is only 1.5\%. This implies that the initial nucleation of oppositely polar domains appears to be energetically more costly in boron containing samples than in pure \Alx - a trend which continues as the boron concentration is  increased.

The shift in positive $E_c$ with cycling follows the logarithmic relationship from the start for both \AlSc and \AlScBx. However, the slope of the fit is significantly steeper for \AlScBx, indicating a more pronounced variation in imprint.

The results are a good indication that the increase in imprint upon boron incorporation is a factor that may explain why wake-up for \AlBx is discussed more frequently in literature than for \Alxs.\cite{Zhu2022WakeUpFilms,Yazawa2023PolarityFerroelectrics,Yazawa2023AnomalouslyModel} As shown in previous work, the formation of inversion domains at the electrodes during cycling enables switching of the polarity at lower coercive fields, stabilizing the non-native polarity and resulting in a shift of the hysteresis towards more symmetric fields during the first few cycles. If the applied electric field amplitude is too small due to limited breakdown resistance, $E_c$ falls outside the measurable range and the film switches partially in the first cycles. Due to the imprint shift, this volume increases in each cycle until $E_c$ lies fully in the measured voltage range and full polarity is reached after several cycles. As discussed above, it is apparently energetically more costly to nucleate the opposite polarity in an as-deposited film when the boron content is increased. This results in a higher negative coercive field for originally N-polar samples and thus a higher chance to observe wake-up - which in our study was most pronounced for \AlBxs without Sc incorporation.




\subsection{Conclusion}

This study investigated the impact of temperature and boron concentration on the structural and ferroelectric properties of \AlScBxs films Higher temperatures lead to improved 0002 RC-FWHM values, indicating better crystallinity. Structural analyses via SEM also show a reduction in AOGs at elevated growth temperatures. 
Boron concentration notably affects the polarity of the films. Films with 6\% to 13\% boron content exhibit N-polar growth, while those with 16\% boron show mixed polarity. This suggests that boron containing films might exhibit a higher tendency for M-polarity.
With respect to the ferroelectric performance, the introduction of boron into \Alxs increases the coercive field and the breakdown strength, respectively; however the $E_c$ increases stronger than $E_{Bf}$. A significant change in ferroelectric properties from \Alxs over \AlScBx to \AlBxs is also evident from the increasing imprint shift with higher boron concentration. While the negative coercive field shifts by 1.5\% in \AlSc in the first two cycles, it shifts by 6\% for \AlScBx with 13 at.\% boron. The underlying tendency to require higher electric fields to switch the initial polarity provides an explanation for the more frequent occurrence of wake-up effects in \AlBxs compared to \Alx. 
For leakage reduction, incorporating boron into \Alx is shown to be beneficial and is correlated to $E_g$. Up to 13\% Boron, the \AlScBx films demonstrate a significant reduction in leakage currents compared to both \Alx and \AlBxs. The reduction of leakage current is crucial for improving the performance and reliability of ferroelectric devices, making \AlScBx films potentially advantageous for applications requiring low leakage currents.

\section{Supplementary Material}
Additional Information on Electrical Measurements, calculations of lattice parameters and supporting Graph for bandgap estimations by Ellipsometry.

\section{acknowledgments}
This collaborative work was enabled through funding by the Federal Ministry of Education and Research (BMBF) under project no. 03VP10842 (VIP+ FeelScreen) and the Deutsche Forschungsgemeinschaft (DFG, German Research Foundation), Project ID 458372836. Funded by the European Union (FIXIT, GA 101135398). Views and opinions expressed are however those of the author(s) only and do not necessarily reflect those of the European Union or the European Research Council Executive Agency. Furthermore, this work was partly supported by SUPREME, one of seven centers in JUMP 2.0, a Semiconductor Research Corporation (SRC) program sponsored by DARPA. The authors acknowledge the use of the Cornell NanoScale Facility (CNF), a member of the National Nanotechnology Coordinated Infrastructure (NNCI), which is supported by the National Science Foundation (NSF) Grant No. NNCI-2025233.”

\section{Data Availability}
The data that support the findings of this study are available from the corresponding authors upon reasonable request. 

\section{References}
\bibliography{references}

\end{document}

%% file: preamble.tex
\usepackage{amsthm}
\usepackage{mathtools}
\usepackage{physics}
\usepackage{xcolor}
\usepackage{graphicx}
\usepackage[left=23mm,right=13mm,top=35mm,columnsep=15pt]{geometry} 
\usepackage{adjustbox}
\usepackage{placeins}
\usepackage[T1]{fontenc}
\usepackage{lipsum}
\usepackage{csquotes}
\usepackage{wrapfig}